\documentclass[12pt]{article}
\usepackage[utf8]{inputenc}
\usepackage{graphicx}
\usepackage{hyperref}
\usepackage[normalem]{ulem}

\DeclareUnicodeCharacter{2212}{-}

\begin{document}

\title{Modeling the Popularity of Twitter Hashtags with Master Equations}
\author{Oscar Fontanelli$^1$, Ricardo Mansilla$^1$ \\
{\small  1. Centro de Investigaciones Interdisciplinarias en Ciencias y Humanidades}\\
{\small   Universidad Nacional Aut\'{o}noma de M\'{e}xico,}\\
{\small  Circuito Exterior, Ciudad Universitaria, M\'{e}xico City,  04510, M\'{e}xico.  }
}
\date{\today}

\maketitle


\begin{center}
{\bf Abstract}
\end{center}

In this work we introduce a model based on master equations to describe the time evolution of the popularity of topics and hashtags on the Twitter social network. Specifically, we model the number of times a certain hashtag appears on the network as a function of time. In our model, the behavior of this quantity depends on the degree distribution of the network and the extrinsic interest the community has for the topic or hashtag. From the master equation, we are able to obtain explicit solutions for the mean and variance. We propose a gamma kernel function to model the topic popularity, which is quite simple and yields reasonable results. Finally, we validate the plausibility of the model by analyzing actual Twitter data obtained through the public API. \\


\section{Introduction}

The emergence and popularization of social networking services constitutes an unprecedented social phenomenon that has transformed the way people communicate, get access to different kinds of information, establish communities and many other things. These novel communication channels allow for the fast and massive diffusion of both information and disinformation, a feature that has been well exploited by marketing agencies, social movements, political parties and government agencies, among others. It is therefore relevant to understand the process of information diffusion over this kind of networks.

Among the most popular social networking services, such as Facebook, YouTube or Instagram, the microblogging site Twitter stands as particularly effective for information diffusion purposes. According to 2016 data (\url{about.twitter.com}), Twitter has approximately 320 million active users (accounts that show activity at least once a month), which represent approximately 9 $\%$ of total Internet users worldwide (\url{www.itu.int}). According to these same sources, approximately 500 million messages are sent over this network everyday. 

The growing interest in modeling and understanding different dynamical processes that occur on this social network is manifested in the large number of studies on this matter in recent years. Kawamoto et al. have proposed a multiplicative process model for information spread \cite{kawamoto2013,kawamoto2014}. Kwon et al. have proposed models for the evolution of the number of messages, the propensity to send or resend messages and have categorized messages according to predictability and sustainability \cite{kwon2012,kwon2013,ko2014}. Weng et al. have elaborated and agent-based model for information overflow and have discovered similarities between images diffusion over Twitter and epidemic spreads \cite{weng2012,weng2013}. Mathiesen et al. have studied scaling laws of big brands tweet-rates, which have been modeled through classic stochastic equations \cite{mathiesen2013,mollgaard2015}. Sutton et al. have made statistical analysis for the diffusion of official warnings during disasters and have identified some factors that contribute to information diffusion \cite{sutton2015}. There are also some works that model topic popularity and information spread with SIR or SIRI-like equations \cite{xiong2012,jin2013,skaza2017}. Bao et al. have studied the predictability of the numer of times a message will be shared or resent \cite{bao2019}. Bauman et al. have modeled community polarization on social networks and specifically analyzed this with Twitter data \cite{baumann2020}. Yook et al. have developed models to account for the observed probability distributions and scaling laws of images and topics popularity \cite{yook2020}. There are as well many other studies for different kinds of phenomena that occur on this social network, other than dynamical process, see for example \cite{gonccalves2011,bovet2018,bovet2019,zhang2018}. Finally, there are many other studies for this kind of phenomena on other social networks, see for example \cite{crane2008,hogg2009,wu2007,miotto2014,miotto2017,wang2018}.

In this paper we propose and validate a model, based on master equations, for the temporal evolution of the number of times a certain topic or label appears on the Twitter network (these labels are called \emph{hashtags}, as we explain in the next section). Notice this is not a model for the number of times a message is sent or shared, but for the number of times it appears on the network, which depends on the number of links the nodes that are sending this message have (the degree distribution of the network). Clearly, a label being shared by nodes with a few links will behave differently, on a global scale, than a label being shared by nodes with many links. We use this as measure of popularity for the label or topic and construct our model under the hypotheses that this popularity is influenced by the degree distribution (a feature that is intrinsic to the network)  and also by the extrinsic popularity of the topic (see \cite{bandari2012} for a discussion on this subject). Data obtained through the Twitter API show that our model is indeed plausible. As far as we know, this is the first attempt to approach this phenomenon with semi-deterministic models.

This paper is organized as follows: in section \ref{section-network} we describe the phenomenon we want to study on Twitter in terms of network theory; in section \ref{section-model} we develop the model, based on master equations; we show in section \ref{section-solutions} how to obtain solutions for the mean number of messages and its variance; in section \ref{section-popularity} we explain how we modeled the extrinsic topic-popularity function; in section \ref{section-validation} we show how we calibrated the model data and demonstrate that the model is consistent with empirical data from Twitter; finally, section \ref{section-discussion} discusses implications and limitations of the model, as well as future research paths. 

\section{Twitter as a directed network}
\label{section-network}

From a network perspective, Twitter is a directed network where nodes are Twitter users and links represent a follower/friend relationship between them. Users interact on the network by sending messages called \emph{tweets}. Not every user on the network receives all messages. A \emph{follower} of user $i$ is a user that receives all messages sent by $i$. If $j$ is a follower of $i$, then $j$ receives messages sent by $i$ but not the other way around. If $j$ is a follower of $i$, then we say that $i$ is a \emph{friend} of $j$, and represent this in the adjacency matrix of the network through $a_{ij}=1$. In this way, there is a directed link in the network from node $i$ to node $j$, through which a message can flow. Every time $i$ sends a message, all of its friends receive it. If a user receives a message and decides to resend it to his of her followers, we say that this user \emph{retweets} the message. We say that the original message is a tweet and the resent message is a \emph{retweet}. In this way, a specific message can propagate through the network via retweets.

A \emph{hashtag} is a keyword or phrase used to describe a certain topic or theme. Hashtags are preceded by the hash sign ($\#$) and they are widely used because they categorize tweets in a way that is easy for other users to find. Many different messages can be categorized by a common hashtag; if this is the case, all these messages usually speak about a common topic or theme. A certain hashtag propagates through the network if users retweet messages that contain it, or if they send new messages categorized by the same hashtag. A hashtag propagates and popularizes when many users are sending messages about a topic of current interest.

A word, phrase, topic or hashtag that is mentioned at a greater rate than others is said to be a \emph{trending topic}. Trending topics become popular either through a concerted effort by users or because of an event that prompts people to talk about a specific topic. We recall that the purpose of this work is to model with master equations the popularity evolution of a hashtag or topic. We develop this model in the next section.  

\section{The model}
\label{section-model}

For simplicity, we assume that users read all messages they receive from their friends immediately after these are sent. Therefore, if a user with $n$ followers sends a message, we say that this message has $n$ \emph{reads} (indicating that $n$ users have received it). We want to model the time evolution for the number of reads $X(t)$ of all messages categorized by a specific hashtag. In this way, $X(t)$ is a measure of the popularity of a certain topic, phrase or news on the network at time $t$. At any fixed time, we consider $X(t)$ to be a random variable; our goal is to find an equation for the probability of having exactly $X$ reads of a certain hashtag at time $t$, which we denote $P(X=x,t)$.

We say that a user \emph{shoots} every time he or she sends or resends a message with the hashtag of interest. Let $N$ be the total number of users in the community. Let $w(t)$ be the average rate at which users shoot. This means that the average probability for every user to shoot in the time interval $(t,t+dt)$ is $w(t)dt$. Finally, let $f(y)$ be the out-degree distribution of the network, so the probability of a randomly picked user to have $y$ followers is $f(y)$. The contributions to $P(X=x,t)$ are the following:

\begin{itemize}
    \item There were $x$ reads at time $t$ and nobody shot (which happens with probability $1-Nw(t)dt$,
    \item there were $x-1$ reads at time $t$ and exactly one user with $y=1$ follower shot (which happens with probability $Nw(t)dtf(1)$),
    $$\vdots$$
    \item there were $0$ messages at time $t$ and exactly one user with $y=x$ followers shot (which happens with probability $Nw(t)dtf(x)$).
\end{itemize}

Since we will consider the limit of very short time intervals, $dt\rightarrow 0$, other possible contributions, such as more than one user shooting during the interval $(t,t+dt)$, do not need to be included, as their contribution will be of higher order in $dt$. Summing up all contributions we get the equation, from the law of total probability,
$$
P(x,t+dt)=\displaystyle
P(x,t)[1-Nw(t)dt]+Nw(t)dt\sum_{i=1}^xP(x-i,t)f(i) + O(dt^2).
$$

Rearranging terms and taking the continuous-time limit $dt\rightarrow 0$ we obtain the partial differential equation for $P(x,t)$,
$$
\displaystyle \frac{\partial P(x,t)}{\partial t} = 
-Nw(t)\left[P(x,t)-\sum_{i=1}^xP(x-i,t)f(i)\right].
$$

We can further approximate the out-degree distribution $f(y)$ to be a continuous distribution 
with support $[m,\infty)$ so there is a minimum of (possibly zero) $m$ followers per user. With this approximation, we get the equation
$$
\displaystyle \frac{\partial P(x,t)}{\partial t} = 
-Nw(t)\left[P(x,t)-\int_m^x P(x-y,t)f(y)dy\right].
$$
After a change of variable and rearranging terms, we finally get the equation
\begin{equation}
\label{master}
    \displaystyle \frac{\partial P(x,t)}{\partial t} = 
    -Nw(t)P(x,t) + Nw(t)\int_0^{x-m}P(y,t)f(x-y)dy.
\end{equation}
This equation, along with the initial condition of zero reads at time $t=0$,
\begin{equation}
\label{delta}
    P(x,0) = \delta(x)
\end{equation}
constitute a master equation for the evolution of the number of reads containing a certain hashtag in the network. In a mean-field framework, $w(t)$ is the probability density of an average user in the network to send o resend a message at time $t$; therefore, this function represents a measure of the popularity that the topic categorized by the hashtag as at time $t$. If the hashtag under consideration is very popular, then it has a high probability of being mentioned in new messages and the messages that contain it have a high probability of being resent. We will refer to this function $w(t)$ as the \emph{hashtag-popularity function}.

\section{Solutions for the mean and variance}
\label{section-solutions}

Explicit solutions for Eq.(\ref{master}) will depend on the forms of the popularity function $w(t)$ and the out-degree or followers distribution $f(y)$ and will be generally not available. However, we can get an equivalent equation for the moment generating function (mgf) of $X(t)$, which we will denote $M_X(s,t)$ and we will be able to utilize it to derive equations for the mean and variance of $X(t)$.

Consider the Laplace transform with respect to $x$,
$$
L_s^{(x)}[g(x)]= \displaystyle \int_0^\infty e^{-sx}g(x)dx.
$$
Direct integration shows that the Laplace transform of the integral on the right-size of  Eq.(\ref{master}) is
$$
\begin{array}{lll}
\displaystyle L_s^{(x)} \left[\int_0^{x-m}P(y,t)f(x-y)dy\right]
& = & \displaystyle \int_0^\infty e^{-sy}P(y,t)dt \int_m^\infty e^{-sy}f(y)dy\\
 & & \\
 & = &  
L_s^{(x)}[P(x,t)]E_f[e^{-sx}].
\end{array}
$$

From the relationship between the moment generating function and the Laplace transform $L_{-s}^{(x)}[P(x,t)] = M_X(s,t)$ we can derive an equation for $M_X(s,t)$ by taking the Laplace transform of Eq.(\ref{master}),
\begin{equation}
     \label{mgf}
     \frac{\partial M_X(s,t)}{\partial t} = N(M_f(s)-1)w(t)M_X(s,t).
\end{equation}
Here, $M_f(s)$ is the mgf of the out-degree or followers distribution $f(y)$. Taking the Laplace transform of the initial condition Eq.(\ref{delta}) we get
\begin{equation}
    M_X(s,0)=1.
\end{equation}
Because of the popularity function $w(t)$, Eq.(\ref{mgf}) will be in general a non-linear differential equation for $M_X(s,t)$ and we cannot give a general explicit solution. We can, however, utilize the fact that the n-th moment of a distribution, if it exists, is given by the n-th derivative of the mgf evaluated at zero,
$$
E[X(t)^n] =\left. \frac{\partial ^n M_X(s,t)}{\partial s^n} \right|_{s=0}.
$$

For $n=1$, we obtain a very simple equation for the expectation of $X(t)$,
$$
\frac{dE[X(t)]}{dt}=Nw(t)\langle f \rangle, \qquad E[X(0)]=0,
$$
where $\langle f \rangle$ is the first moment of the out-degree distribution, i.e. the mean number of followers of users in the community. This equation has the solution
\begin{equation}
    \label{expectation}
    \displaystyle E[X(t)]=N\langle f \rangle \int_0^t w(s)ds.
\end{equation}
In a similar way, we can get an initial value problem for the second moment,
$$
\frac{dE[X^2(t)]}{dt}=Nw(t)[2\langle f \rangle E[X(t)]+\langle f^2 \rangle ], \qquad
E[X^2(0)]=0,
$$
where $\langle f^2 \rangle$ is the second moment of the followers distribution. Thus,
$$
E[X^2(t)]= \displaystyle N \int_0^t w(s)[2\langle f \rangle E[X(s)]+\langle f^2 \rangle ]ds.
$$
Finally, we can have an expression for the variance of $X(t)$, 
$$
    Var[X(t)]= \displaystyle N \int_0^t w(s)\left[2E[X(t)]\langle f \rangle + \langle f^2 \rangle \right]ds - \left(E[X(t)]\right)^2.
$$

Integrating by parts the first term of the variance, rearranging terms and simplifying, we get
\begin{equation}
\label{variance}
    Var[X(t)] = \displaystyle N \langle f^2 \rangle \int_0^t w(s)ds = \frac{\langle f^2 \rangle}{\langle f \rangle} E[X(t)].
\end{equation}

\section{Modeling the popularity function}
\label{section-popularity}

Consider the simplest possible case, where the interest a hashtag produces remains constant over time, thus $w(t)$ is a constant function. Recall that $w(t)$ is a probability for any fixed time, so it must always lie in the interval $[0,1]$. By using $w(t)=c$ with $c\in [0,1]$, we obtain from Eqs.(\ref{expectation}) and (\ref{variance})
$$
E[X(t)] = Nc\langle f \rangle t, \qquad Var[X(t)]=Nc\langle f^2 \rangle t.
$$

A more realistic consideration is that the interest grows until it reaches a peak, then decays and vanishes for very large times. This behavior can be represented in several ways. Here we will examine one simple possibility, which is a function proportional to a gamma distribution kernel,
\begin{equation}
\label{w}
w(t) = c \frac{e^a}{(ab)^a}t^a e^{-t/b},
\end{equation}
where $a,b>0$ are parameters that control the shape of the interest function and $c\in[0,1]$ is the value of $w(t)$ at its peak. Notice that $w(t)$ reaches its maximum value $w_{max}=c$ at $t_{max}=a\cdot b$ and has an inflection point at $t_{inf}=a\cdot b + b\sqrt{a}$. With this popularity function, we get from Eqs.(\ref{expectation}) and (\ref{variance})
\begin{equation}
    \begin{array} {c}
    \displaystyle E[X(t)] = \frac{Ncbe^a \langle f \rangle}{a^a}\gamma (t/b,a+1), \\
    \\
    \displaystyle Var[X(t)] = \frac{Ncbe^a \langle f^2 \rangle}{a^a}\gamma (t/b,a+1).
    \end{array}
\end{equation}

Here, $\gamma (x,s)$ is the lower incomplete gamma function, $\gamma (x,s) = \int_0^x e^{-t}t^{s-1}ds$. By utilizing the Stirling approximation for the gamma function
$$
\Gamma (z) = \sqrt{\frac{2\pi}{z}}\left(\frac{z}{e}\right)^z \left(1+O\left(\frac{1}{z}\right)\right),
$$
we can approximate the limits for the expectation and variance for very large times,
$$
\begin{array}{c}
\displaystyle E[X(t)] \longrightarrow \frac{Ncbe^a \langle f \rangle}{a^a} \Gamma (a+1) \simeq Nbc\langle f \rangle \sqrt{2\pi(a+1)},\\
 \\
\displaystyle Var[X(t)] \longrightarrow \frac{Ncbe^a \langle f^2 \rangle}{a^a} \Gamma (a+1) \simeq Nbc\langle f^2 \rangle \sqrt{2\pi(a+1)}\\ 
\end{array}
$$
for large values of the parameter $a$.\\

Notice that this is not the only way in which we can model the popularity function, but it constitutes a relatively simple function that yields acceptable fits, as we will see in the following section.

\section{Model calibration and validation}
\label{section-validation}

In order to corroborate the validity of the model, we analyzed time evolution of popular trends and hashtags in Twitter during the first half of February 2020. We obtained data through the Twitter API with the \emph{rtweet} library for the statistical software R \cite{rtweet-package}. We implemented the following pipeline to contrast empirical observations with model predictions:
\begin{enumerate}
    \item From the sample of tweets, directly compute number of different users $N$, mean number of followers $\langle f \rangle$ and mean square number of followers $\langle f^2 \rangle$.
    \item Divide time interval of the sample into $n$ equal length sub-intervals, then compute fraction of different users that sent a message within each sub-interval. This gives us the empirical popularity function $w(t)$, since it approximates the probability for each user to send a message at any time.
    \item Empirical $w(t)$ is usually very noisy, so we smooth this time series with a simple $k$−
    point moving average filter. This gives us a smoothed empirical popularity function.
    \item Fit parameters for theoretical $w(t)$ with Levenberg-Marquardt non-linear least squares.
    \item Utilize the cumulative sum of followers as an empirical approximation for the time evolution of $X(t)$.
    \item With fitted $w(t)$ parameters, and knowing theoretical $E[X(t)]$ and $Var[X(t)]$,
    construct $95 \%$ approximate confidence regions for $X(t)$ and contrast with empiric observations.
\end{enumerate}

\begin{figure}[h]
\includegraphics[width=0.99\linewidth]{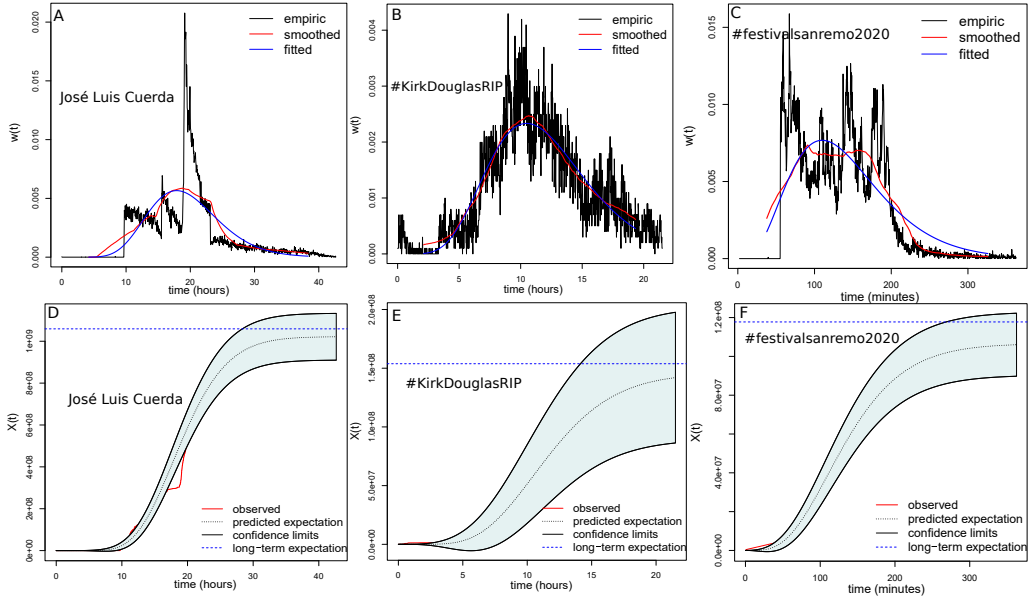}
\caption{ A, B and C) Empiric, smoothed and fitted popularity function $w(t)$, D, E and F) Observed evolution of reads $X(t)$ and approximate confidence region predicted by the model for three different trends and hashtags. Last three figures also show the limit expectation for very large times.}
\label{results}
\end{figure}

We show in Fig. \ref{results}, the results of our analyses for three different trends and hashtags. From the database we collected, we chose three worldwide trending topics on three different time scales: first, the trending topic ``\emph{José Luis Cuerda}'', following the decease of this Spanish film director on February 4, 2020; this was a world trend for approximately two days. Second, the hashtag $\#KirkDouglasRIP$, which was a world trend for approximately one day after the decease of the American actor and film producer on February 5, 2020. Third, the hashtag $\#festivalsanremo2020$, which was a world trend for approximately three hours during the grand final of the San Remo Music Festival 2020 on February 8, 2020. The first three panels on this figure show the empiric popularity function, computed directly from the Twitter data, as well as the smoothed and fitted popularity functions. Notice how the empiric $w(t)$ is somewhat noisy, yet fitted and smoothed functions are very close to each other. The last three panels show in red the empiric number of reads $X(t)$, dashed lines are the expected number of reads predicted by the model, we show in shaded blue the approximate confidence region for $X(t)$ and the dotted blue line is the long-term expectation predicted by the model. Notice how in these tree cases the observed $X(t)$ stays within the confidence region for almost the entire time intervals. See, for example, how wide the confidence region is for $\#KirkDouglasRIP$ in comparison with the other two, which is a consequence of a relatively larger variance on the degree distribution for this community. 

\section{Discussion and conclusions}
\label{section-discussion}

We have presented a mathematical model, based on master equations, for the temporal evolution of the popularity of a certain hashtag or topic on the Twitter network.  The measure we utilize for the popularity of a hashtag is the number of times it appears on the network, which depends on how many users have posted it and how many followers these users have. According to our model, there are two main components that influence this dynamics: on one side, certain characteristics of the community and the network such as number of nodes and mean and variance of the degree distribution; these are components that are intrinsic to the network. On the other side, we have the time evolution of the interest people have on the topic or hashtag we are modeling, which we quantify as the probability each user in the community has of sending a message as a function of time. This popularity function is an extrinsic component influencing this dynamics.

We utilized actual Twitter data, that we got from the public API, to calibrate the model (fit parameters from the empiric popularity function) and to compare its predictions and the actual observations. Even though we are not able to give an explicit solution for the master equation, we can compute the mean and variance and therefore construct approximate confidence regions. The examples we show in this paper confirm that our model is plausible and consistent with the observations. 

We have used only one possibility to model the popularity function, one that is relatively simple and yields acceptable fits. However, other functions with similar behaviors may be used. More important than this is the fact that the parameters of this function are fixed, ignoring the possibility that the shape of the popularity function varies with time, for example through a back-feeding  process (a popular hashtag gets more and more popular over time). The possibility of a popularity function that updates and that is itself an unknown function is a matter of future study.

The Twitter public API we utilized to gather our data base has some limitations: we can only make 18 thousand requests every 15 minutes and we can only access tweets that are 10 days old or newer. We believe that a more comprehensive data base would be helpful and illustrating to see the performance of our model on a more global scale. In spite of these limitations, we observed that our model is consistent with the observations. This is also a matter of future study. 

Accurately predicting the evolution and impact a certain tweet or hashtag will have on the network is a difficult task and it is currently a matter of great interest. With this model, we hope to contribute to the understanding of this phenomenon. Finally, the activity on Twitter may not be completely different from dynamics on other social networks, online or offline; we believe that the present model, though very simple, can give interesting insights into the behavior of other networks.\\

\noindent \textbf{Author contributions:} Conceptualization, data collection, methodology, visualizations, writing, reviw and editing by OF and RM.

\noindent \textbf{Competing interests:} The authors declare that they have no known competing financial interests or personal relationships that could have appeared to influence the work reported in this paper.

\noindent \textbf{Acknowledgments:} Suggestions and comments by Mario Alejandro López Pérez and Ricardo Mansilla Sánchez are gratefully acknowledged. This study was supported by the UNAM-DGAPA Postdoctoral Scolarships Program at CEIICH, UNAM. 

\bibliographystyle{unsrt}
\bibliography{references}

\end{document}